\documentclass[suppldata]{interact}

\usepackage{blindtext}
\usepackage{booktabs}      % For professional looking tables
\usepackage{tabularx}      % For tables with text wrapping
\usepackage{geometry}
\usepackage{enumitem}
\usepackage[table]{xcolor}
\usepackage{multirow}
\usepackage{array}
\usepackage{graphicx}      % for \rotatebox, subcaption, etc.
\usepackage{subcaption}
\usepackage[utf8]{inputenc}
\usepackage{titlesec}
\usepackage{float}

\titleformat{\paragraph}[runin]
  {\normalfont\normalsize\itshape}
  {\theparagraph}
  {1em}
  {}

\setlist[itemize]{noitemsep, topsep=0pt, left=1.5ex}
\setlist[enumerate]{noitemsep, topsep=0pt, left=1.5ex}

\definecolor{headerblue}{HTML}{2F66B2} % deep blue
\definecolor{cellblue}{HTML}{DCE9FB}   % pale blue

\newcolumntype{Y}{>{\raggedright\arraybackslash}X}
\newcolumntype{L}[1]{>{\raggedright\arraybackslash}m{#1}}

\usepackage[natbibapa,nodoi]{apacite}
\usepackage[colorlinks=true, linkcolor=blue, citecolor=blue, urlcolor=blue]{hyperref}

\begin{document}

\title{\textbf{Audio Video Verbal Analysis (AVVA) for Capturing Classroom Dialogues}}

% --- T&F native author formatting (Replaces authblk) ---
\author{
\name{Vivek Upadhyay\textsuperscript{a}\thanks{CONTACT Vivek Upadhyay Email: viveku@iisc.ac.in} and Amaresh Chakrabarti\textsuperscript{a}}
\affil{\textsuperscript{a}Department of Design and Manufacturing, Indian Institute of Science, Bengaluru, India}
}

\maketitle

\begin{abstract}

\textbf{Background:} The classroom discourse analysis has been transformed by the growing use of audio-video multimodal data, which demands analytical methods that balance interpretive depth with computational scalability.

\textbf{Methods:} This study introduces the Audio Video Verbal Analysis (AVVA) framework, 
adapted from the Verbal Analysis method to integrate qualitative interpretation with 
quantitative modelling. Unlike fully multimodal learning analytics approaches, AVVA 
focuses on verbatim transcripts with essential interactional modalities. 

\textbf{Findings:} The framework embeds triangulation as a core design strategy across ten methodological steps, strengthening validity and analytical rigour. A comprehensive validation scheme addresses fundamental challenges in temporal observational research: Phi Ceiling for low-frequency variables (via Base-Rate Filtering), estimation uncertainty (via bootstrap confidence intervals), and the Modifiable Temporal Unit Problem, where measured associations depend on observational window size. Four-criterion stability assessment (sign consistency, confidence interval overlap, zero exclusion, magnitude stability) classifies variable pairs into interpretable patterns: grain-invariant, scale-specific, or multi-scale, etc. structures across temporal grain sizes. Its application to 23 hours of classroom recordings illustrates its practical viability and its potential to yield meaningful insights.

\textbf{Contribution:} The framework thus provides a scalable pathway for transforming rich classroom discourse into analysable datasets.

\end{abstract}


\section{Introduction}
In the social sciences, video data have been extensively employed to examine teaching and learning environments \citep{Jewitt2012, Lackovi2018, Ramey2016VideoAnalysis, ROSENSTEIN2000373, Rosenstein2002, Walker2018}, courtroom interactions \citep{Bannon2020}, and ethnographic settings \citep{Hackett2015-bv}. Research within the medical sciences has predominantly utilised video-based observation for educational and training purposes, such as studies of doctor–patient communication \citep{Kazmi2014, Pearce2010}, the instruction of medical personnel \citep{Topor2016}, and investigations concerning surgical training and patient safety practices \citep{Guerlain2004, Topor2016}. Thus, video-based research enables us to move beyond verbal interactions, allowing the analysis of multiple dimensions of human behaviour and social activity across diverse contexts \citep{Ramey2016VideoAnalysis}. Video provides a deeper level of analysis due to the diversity of the data it captures. It enables the examination of interactions, simultaneous actions \citep{Norris2004}, and bodily movements \citep{Bezemer2008}. Therefore, audio–video based analysis is often regarded as a valuable enhancement to traditional transcript-based analysis, offering a more accurate account of events as they occurred \citep{Merriam1998QualitativeRA} while preserving the expressive and contextual richness of communication \citep{Crichton2005}. Although using audio and video data can make analysis more complex at first, relying only on written transcripts may lead to a loss of important context in participants’ narratives \citep{Crichton2005, Schnettler}.\\

The growing recognition of the analytical potential of audio-video data underscores the need for a methodology that connects qualitative interpretation and quantitative analysis. Researchers typically begin with qualitative exploration, identifying impressions and emergent patterns, and then develop coding schemes to formalise these observations. Once coded, the data can be examined quantitatively, effectively transforming subjective impressions into operationalised, measurable constructs through the comparative analysis of code frequencies \citep{chi1997}. To address this need, \citep{chi1997}  proposed Verbal Analysis, a methodological approach that offers systematic guidance for analysing verbal data by integrating qualitative interpretation with quantitative rigour, thereby reducing subjectivity in interpretation. While audio–video recordings capture the multimodal richness of human interaction, they also introduce considerable complexity to analysis \citep{Bateman2008}. In response, the present study introduces the Audio Video Verbal Analysis (AVVA) framework, designed to enable scalable, systematic verbal analysis alongside large audio-video datasets. The ten steps of the AVVA framework emphasise both methodological robustness and practical feasibility, with provisions for extending the encoded data to applications in machine learning and deep learning. Across these steps, various types of triangulation \citep{denzin1978, patton1999} have been implemented as a design strategy. During the statistical validation of patterns, it was realised that variables with low frequencies of occurrence yield unreliable correlation measures \citep{Bakeman1997}; hence, the concept of a base-rate filter has been introduced. Bootstrap confidence interval analysis \citep{Efron1993} quantifies estimation uncertainty, enabling researchers to differentiate patterns from sampling variability. During the grain-size experiments, challenges related to temporal aggregation, such as the Modifiable Temporal Unit Problem (MTUP) \citep{cheng2014modifiable}, have been observed. MTUP describes how observed relationships vary with the chosen time window, and is analysed using four criteria to assess stability. Variable pairs are classified into interpretable patterns (grain-invariant, scale-specific, multi-scale, etc.) based on sign consistency, confidence interval overlap, zero exclusion, and magnitude stability across multiple temporal grain sizes. The AVVA framework was applied to collect approximately 3 months of audio-video data. Then, following all the steps of the AVVA framework, we encoded 23 hours of classroom recordings from science and mathematics lessons across grades 6–12 in an Indian school, for the variables Knowledge Dimension, Cognitive Process Dimension, Learning Theories, and 21st-Century Skills.

The following section reviews the most common theoretical frameworks for gathering and analysing visual data and verbal analysis, evaluating their strengths and limitations.
 
\section{\raggedright Analysis of audio video data: An overview of relevant theories}
Actor Network Theory, Picture Theory, Grounded Theory and Multimodality Theory offer insights into visual data analysis \citep{Fazeli2023}. However, not all of them focus specifically on video data or provide clear guidelines on how to organise, describe and interpret video content \citep{Fazeli2023}. Actor-Network Theory (ANT) foregrounds the interplay between human and non-human entities within intricate networks, emphasising the complexity of their interrelations. However, it has been criticised for its conceptual ambiguity regarding how “actors” are to be defined, interpreted, and empirically examined \citep{Williams1996, Wacker1998, Law1999ActorNetwork, Kaghan2001}. Similarly, Picture Theory advances the primacy of visual representation over linguistic expression \citep{Mitchell1995-MITPTE-2}, offering valuable insights into visual analytical methodologies. Nevertheless, as its focus remains largely on static imagery, its application to dynamic video analysis remains limited.

Grounded Theory derives theoretical insights directly from empirical data rather than from pre-existing theoretical frameworks, hence its name. Originally formulated by \citep{Glaser1967}, subsequent scholars extended the approach to encompass a postmodern dimension \citep{Charmaz2006ConstructingGT, Clarke2017-bw}. The methodology is grounded in the assumption that reality consists of multiple constructions, with the researcher inherently connected to the knowledge being produced. This perspective supports a flexible set of principles and practices for theorising about how individuals interpret and construct their lived experiences \citep{Charmaz2014-pw}. Grounded Theory has had a significant impact across disciplines, including psychology, nursing, medicine, social work, the social sciences, and education \citep{Holton2017, Timmermans2012, Creswell2017-qh}.

Originating from the work of Gunther Kress, Multimodality Theory provides a critical framework for investigating social communication at the intersection of linguistic and visual modes \citep{Hodge1988-xz, Kress2010}. The theory explores how meaning is produced and interpreted through diverse communicative modes, such as speech, gesture, gaze, and visual imagery, beyond traditional written language \citep{KressVanLeeuwen2001}. Owing to its robust foundations for analysing both visual and non-visual semiotic resources, Multimodality Theory has been widely applied to the study of communication through still and moving images, video, audio, and three-dimensional artefacts \citep{Dicks2020-aw}. It offers a comprehensive analytical lens that encompasses a spectrum of meaning-making practices, from textual and gestural expressions to the use of colour, typography, and visual composition, particularly relevant in an era where technological advancements continually expand the modalities through which people interact \citep{Dicks2020-aw, Kress2010}. There are, however, disadvantages to Multimodality Theory, including the multimodal approach may be overly time-consuming and lengthy, hence inadequate for managing large data sets, while leading to unjustified conclusions or conclusions based on limited samples \citep{Bateman2008}. In education, the emerging field of Multimodal Learning Analytics (MMLA) aims to integrate data from students' speech, eye gaze, posture, and physiological signals to create a holistic representation of learning processes, though significant challenges in data synchronisation, model interpretability, model opacity, and ethical considerations surrounding privacy and consent. Furthermore, there is a need to improve the scalability and reproducibility validity of multimodal systems, especially in underrepresented educational settings \citep{GuerreroSosa2025}.  

During the 1990s, the growing accessibility of digital video recording technologies and computer-assisted qualitative data analysis software CAQDAS such as ATLAS.ti, MAXQDA, NVivo, and Dedoose expanded the methodological toolkit available to researchers. This technological advancement facilitated alternatives to conventional text-based coding, enabling the direct analysis of audio and video data segments \citep{Evers, Bezemer2011, Bassett}. With the proliferation of visual digital technologies, including video cameras, smartphones, and computers capable of creating and displaying video \citep{ChawlaDuggan2019}, research employing video-based methods has continued to grow \citep{Bezemer2011}. However, as \citep{Pink2013-mj} cautions, researchers must critically reflect on the epistemological and methodological implications of these technologies to ensure analytical rigour and validity. The integration of visual methods enriches data interpretation by providing deeper insights into the phenomena under investigation. Particularly in research involving children, video methods enable a more nuanced understanding of participants’ perspectives, often capturing experiential dimensions that textual data alone may fail to convey \citep{Craig2021}.

Beyond their analytical utility, audio and video data also hold significant potential for knowledge translation. Incorporating such media into scholarly and community dissemination, for instance, through conference presentations or digital publications, can enhance engagement and understanding \citep{Friend2014}. However, despite these advantages, there remains limited clarity regarding the analytical and technical procedures associated with video-based research and the diverse applications of CAQDAS tools \citep{Rahman2016, Silver, fielding1998computer, Bezemer2011}. Consequently, the methodological gap in the literature on qualitative data analysis continues to expand. Moreover, integrating CAQDAS into qualitative research presents several challenges. The steep learning curve of these tools poses difficulties for researchers with limited technical proficiency, and delegating such tasks to research assistants is not always practical \citep{Rahman2016, Silver}. Institutional constraints further exacerbate this issue, as some universities lack the financial or infrastructural capacity to support software procurement \citep{Atieno, fielding1998computer}. Additionally, restrictions in software licensing often hinder cross-institutional collaboration \citep{Silver}.

Verbal analysis is a methodological approach that seeks to quantify the qualitative coding of verbal utterances by systematically tabulating, counting, and relating different types of statements to minimise subjectivity in interpretation \citep{chi1997}. It has been employed to categorise learners’ explanations, such as identifying whether an utterance constitutes an inference, a monitoring statement, or an unrelated remark, thereby providing insight into the cognitive processes underlying comprehension \citep{Chi1994}. Although originally grounded in cognitive research aimed at understanding knowledge acquisition, the method can be extended to examine non-cognitive domains (e.g., social, motivational, or behavioural) and adapted for use with observational or video data. While \citeauthor{chi1997}'s (\citeyear{chi1997}) verbal analysis provides a systematic method for quantifying qualitative data, several scholars have noted inherent limitations in its application. The method presupposes that verbal data have already been collected and transcribed, omitting the methodological complexities of data capture and transcription itself \citep{loubere2017reflexive, heath2010video}. Consequently, it remains largely silent on how analytic validity may be affected by transcription choices or by non-verbal modalities such as gesture and video, which require distinct interpretive frameworks \citep{jordan1995interaction, goldin1993gesture, goldin2012gesture}. Furthermore, verbal analysis is highly time-intensive and resource-demanding, posing significant challenges for scalability and reproducibility in large datasets \citep{miles1994qualitative, saldana2013coding, nassauer2019video}. Critics have also highlighted the interpretive subjectivity inherent in protocol and verbal report methods, noting the absence of standardised procedures and risks of over-interpretation \citep{austin1998protocol, luque2002protocol}. Finally, the emergence of multimodal approaches underscores that purely verbal methods often neglect important non-verbal and contextual cues in meaning-making \citep{jewitt2013multimodal}. Together, these critiques suggest that while \citeauthor{chi1997}'s (\citeyear{chi1997}) framework offers a rigorous foundation for coding verbal data, its extension to dynamic, multimodal, or large-scale contexts remains methodologically and practically constrained.

\section{\raggedright Audio Video Verbal Analysis (AVVA): A comprehensive analytical method}
Taking into account the scalability, time consumption, multimodality, applicability, and triangulation, we have developed a method to facilitate verbal analysis \citep{chi1997} in conjunction with audio and video data. We are providing a systematic, layered, mixed-method approach (components of quantitative and qualitative analysis are both present), which is easy to follow. The AVVA method follows the same core structure as other qualitative approaches \citep{Braun01012006, Elo2008-ks} for data collection, extraction, coding, and reporting (inductive design) in conjunction with the verbal analysis method \citep{chi1997}. The AVVA framework includes the following steps: 
\begin{enumerate}
    \item Data collection; 
    \item Sampling the data;
    \item Transcribing the data;
    \item Selecting units of analysis;
    \item Developing a coding scheme; 
    \item Developing operational coding scheme;
    \item Validity and reliability; 
    \item Representing the coded data;
    \item Finding patterns and coherence in the represented Data;
    \item Interpreting the pattern and its validity
\end{enumerate}
\vspace{1em}
This framework is also rich in triangulation as a design strategy. Triangulation involves the use of multiple research methodologies to study a phenomenon, enabling a more comprehensive and in-depth understanding \citep{denzin2000handbook, meijer2002multimethod, Schaap2011}. \citet{denzin1978} and \citet{patton1999} identified four types of triangulation: (a) method triangulation, (b) investigator triangulation, (c) theory triangulation, and (d) data source triangulation. We operationalise four types of triangulation systematically: (a) data triangulation (Steps 1–2) through collection across temporal, spatial, and contextual dimensions;

\begin{table}[H]
\centering
\caption{Ten steps for analysing audio-video data using the AVVA framework}
\label{tab: AVVA}
\renewcommand{\arraystretch}{1.2} % increases row spacing for readability
\begin{tabularx}{\textwidth}{@{}p{0.8cm} p{4.2cm} X@{}} 
\toprule
\textbf{Step No.} & \textbf{Steps} & \textbf{Description} \\ 
\midrule
1 & \textbf{Data collection} &
Video recording in natural settings, Ethical compliance, and Data triangulation \\[0.6em]

2 & \textbf{Sampling the data} &
Probability sampling, Non-probability sampling, Select sampling strategy suited to research goals, and Data triangulation \\[0.6em]

3 & \textbf{Transcribing the data} &
Manual or automated transcription, Time-stamping, Naturalised vs. denaturalised transcription styles, Inclusion of multimodal cues, Flexibility to revisit transcripts \\[0.6em]

4 & \textbf{Selecting units of analysis} &
Granularity, Correspondence, and Segmentation guided by non-content or activity-based cues \\[0.6em]

5 & \textbf{Developing coding schemes} &
Inductive (data-driven) vs. deductive (theory-driven) coding approaches, Conventional content analysis, Directed content analysis, Summative content analysis, and Theoretical triangulation \\[0.6em]

6 & \textbf{Developing operational coding schemes} &
Refining codes for application, Resolving ambiguity and contextual dependence, Code refinement with experts, Triangulating verbal and nonverbal data sources, Chunking, Capturing
macro-context through global variables, and Theoretical triangulation \\

7 & \textbf{Validation and reliability} &
Validation through expert review of operational codes, IRR through both per cent agreement and Fleiss’ kappa, Two-phase training for coders, and Investigator triangulation\\

8 & \textbf{Representing the coded data} &
Bar plots of frequency vs variable and time spent vs variable, Pie charts, etc., Time plots, semantic networks, Argument chains, State transition diagrams, and Methodological triangulation\\

9 & \textbf{Finding patterns and coherence in the represented data} &
Statistical association via parametric and non-parametric approaches, Interaction metrics, Unsupervised learning (clustering, PCA, LSA, etc.) Machine learning, Deep learning (word embeddings, NLP techniques), Computational efficiency and scalability, and Methodological triangulation \\

10 & \textbf{Interpreting the pattern and its validity} &
Triangulation of perspectives, Statistical validation (Grain-Size Experiments, MTUP, Phi Ceiling and Base-Rate Filtering, Bootstrapping, Stability assessment) \\
\bottomrule
\end{tabularx}
\end{table}
(b) theoretical triangulation (Steps 5, 6, 10) through multiple conceptual frameworks in coding development and interpretation; (c) investigator triangulation (Step 7) through inter-rater reliability and collaborative validation; and (d) methodological triangulation (Steps 8–9) through multiple analytical and representational approaches. This systematic integration of triangulation strategies enhances the validity, reliability, and comprehensiveness of findings. Table~\ref{tab: AVVA} presents the ten steps of the AVVA framework, with brief descriptions and relevant triangulation types.

\subsection{Data collection}
Video recordings from natural settings can be used to capture rich and comprehensive data. The video data helps in bringing down the dependency on the selectivity of the observer \citep{chi1997}. Also, collecting data in multiple modalities (audio and video) for continuous months embeds data triangulation. The video data can be collected using a simple camera or a set of multiple cameras. A lightweight camera and tripod arrangement that is portable and easy to transport is essential if a researcher needs to visit many classrooms to get data from various topics. Recent advancements in mobile camera technology have made it possible to utilise a mobile camera with a suitable and adaptable stand. A multiple-camera configuration with a wide-angle lens and a standard lens on a tripod can effectively capture videos if all activities are confined to 1 or 2 classrooms. The researcher should ensure that both video and audio are of good quality. It is advisable to use a stationary, wide-angle lens camera, as it can record clear footage while minimally disturbing the natural environment. In general, simple camera setups are preferred for such recordings. Concerns have been expressed about the invasive nature of video cameras and their possible influence on conduct, which undermines the legitimacy of video data \citep{Schuck}. Additionally, other studies have shown that participants are typically too engrossed in their activities, such as teaching a class of children, to modify their behaviour due to the presence of a video camera \citep{Gobo2008, Pink2007}. To address these issues, any device used for data collection, especially in teaching environments, should be as unobtrusive as possible. Adequate time, typically around one to two months, should be allotted to help participants become familiar with the presence of the camera. Research has shown that with regular exposure, participants gradually adapt to the camera’s presence, and its intrusive effects diminish as they become more comfortable and engaged in the research process \citep{Fitzgerald2013-lo, Jordan01011995}. Since video data occupies substantial hard drive space, its storage and organisation become critical parts of the data collection process. To prevent data loss, multiple copies should be maintained—either on external hard drives or by uploading them to reliable cloud storage services. Videos can be systematically organised by using simple and meaningful file names that include details such as the start time of recording, date of collection, and the topics covered. Additionally, while collecting and later using video data for analysis or discussion, it is essential to address all ethical and privacy considerations. This includes obtaining consent forms from participants, securing permission letters from the head of the institution, and ensuring privacy through methods such as masking participants’ identities in the videos. These measures confirm that the collected data will be used strictly for academic purposes. In the implementation of the AVVA technique, we collected continuous, 3 months of data from an Indian school using the above recommendations.

\subsection{Sampling the data}
Regardless of the data-collecting methods employed, it is impossible to compile an exhaustive description of any one environment. Consequently, choices made about sampling must be taken \citep{Erickson1992}. Sampling methods are classified into two primary categories: Probability sampling, which includes simple, stratified, and cluster random sampling, and Non-probability sampling, which encompasses convenience, judgemental, and snowball sampling \citep{Elfil2017-hn}. Researchers must select an acceptable sampling strategy for their investigation.  As noted earlier, sampling should be carried out using data collected after the initial trial period of 1–2 months. When sampling is done over people, space and context, the type of triangulation is data triangulation. For example, in the implementation of the sampling step, we sampled data over grades 6-12, for science and mathematics topics, for 48 episodes.

\subsection{Transcribing the data}
Following the selection of samples, the transcription of the video data is conducted. Transcription is widely recognised as a labour-intensive and time-consuming process, often requiring between three and more than eight hours of work to accurately transcribe a single hour of audio, depending on the transcriber’s typing proficiency \citep{McMullin2021TranscriptionAQ}. Although a tool named Idea-Transcribe has been developed to cut manual transcription time by nearly half while offering several key features preferred by researchers, such as automatic time-stamping and an integrated typing environment \citep{Sarkar}. Nonetheless, manually transcribing large video data (More than 20 hours) still takes a lot of time. Transcription is not merely a mechanical act of converting speech into text; the resulting written document does not represent an entirely objective account of the spoken event. As Davidson (2009)  notes, written language inherently differs from spoken language in syntax, vocabulary, and grammatical conventions. Consequently, the transcriber must make interpretive choices throughout the process, deciding what to include or exclude, and whether to correct errors, repetitions, or grammatical inconsistencies \citep{Davidson}. Bucholtz (2000) characterises this range of transcription practices as a continuum between naturalised (or intelligent verbatim) transcription, which adapts spoken language to written norms, and denaturalised (or full verbatim) transcription, which retains all elements of speech, including disfluencies, errors, and repetitions \citep{BUCHOLTZ20001439}.

Verbal analysis does not encompass the initial stages of data collection or transcription of verbal protocols. Its focus lies not on the methods of data gathering or transcription, but rather on analysing data that has already been collected and transcribed. The specific type of verbal protocols to be obtained is determined by the underlying theoretical questions and hypotheses guiding the research \citep{chi1997}. Whereas the video reveals several interactional modalities; individuals utilise speech, gestures, gaze, body posture, facial expressions, movement, and physical items to communicate ideas and information \citep{Goodwin2013, Hall1999}. \citep{Goodwin2003d} identified a “recursive interplay between analysis and methods of description” (p. 161), wherein the researcher repeatedly observes, re-examines, and documents video-recorded interactions from multiple perspectives to gradually construct and refine an argument. Consequently, a single video recording can yield multiple equally valid transcripts, each shaped by differing research questions, analytical frameworks, and focal phenomena \citep{Ramey2016VideoAnalysis}.

\begin{figure}[h] % 'h' means place it here
    \centering
    \includegraphics[width=0.9\textwidth]{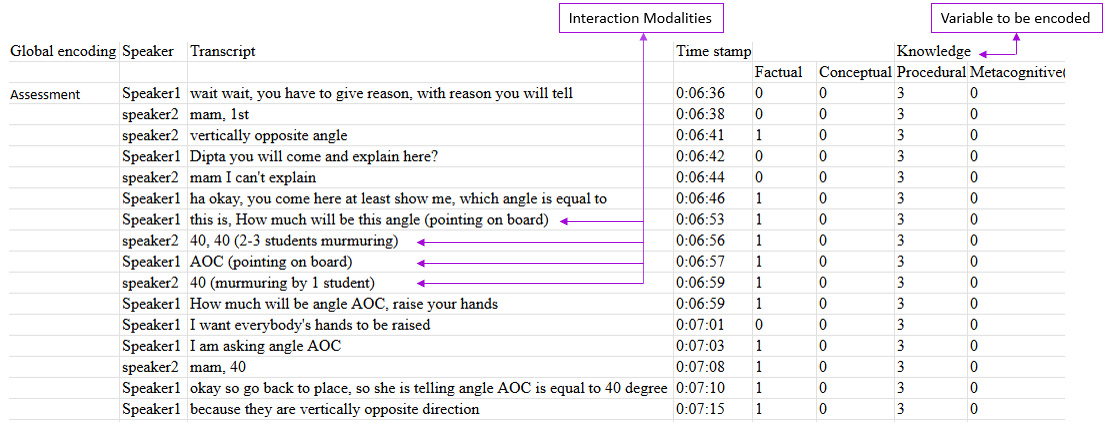} % Replace with your image file
    \caption{Example of the transcription showing interaction modalities}
    \label{fig:1}
\end{figure}

In the AVVA approach, our aim is to perform verbal analysis in conjunction with video data. As mentioned earlier, video data brings opportunities to transcribe various interactional modalities (non-verbal cues). By examining multimodal discourse, which encompasses both verbal and non-verbal signs, researchers can uncover additional levels of meaning, including intention and emotion, in educational encounters \citep{Vivante}. Furthermore, the transcriber must determine the extent to which contextual elements, such as interruptions, overlapping speech, and inaudible sections, should be represented in the transcript \citep{Lapadat01012000}. Given the diversity of qualitative research approaches, there are no universally fixed guidelines for transcription; instead, such decisions should be guided by the specific research questions and methodological orientation of the study \citep{McMullin2023}. Additionally, real-world implementation and scalability remain limited, as much of the existing research in multimodal learning analytics is still conducted within controlled or small-scale environments \citep{GuerreroSosa2025}. As discussed above, transcription is inherently a laborious task, and incorporating multiple interactional modalities, though it enhances the richness of the data, but also introduces challenges related to scalability. To balance detail and feasibility, it is recommended that the transcription include only essential interactional modalities, while subsequent stages of the AVVA approach be conducted alongside the video data as shown in Figure~\ref{fig:1}. Furthermore, although the research questions may be well-defined at the outset, the process of developing a coding scheme may prompt the inclusion of new verbal cues that were not initially captured in the first transcription draft. Since many interactional cues are already preserved within the video recordings, researchers should approach the initial transcription phase with flexibility, remaining open to revisiting and refining specific sections to incorporate additional non-verbal elements as needed during later stages of analysis. As far as triangulation concerns, if multimodal cues like verbal data (what is said), nonverbal data (how it's expressed), and contextual data (physical positioning, movements) have been captured, then it is data triangulation. However, as discussed earlier, because the AVVA technique recommends capturing only essential interactional modalities, data triangulation is applied minimally during the implementation of this framework.

\subsection{Selecting units of analysis}
The next step involves segmenting the verbal utterances to establish the appropriate unit of analysis. Four key factors should be considered during segmentation: (a) the granularity or size of each segment, (b) how this granularity aligns with the research questions, (c) the specific features in the data used to guide segmentation, and (d) instances in which segmentation may not be required \citep{chi1997}.

\subsubsection{Granularity}
For verbal data, the defining cut can occur at many points, revealing units of varying grain sizes, such as a proposition, a sentence, an idea, a reasoning chain, a paragraph, an interchange as in conversational dialogue, or an episode (such as an event, or a specific activity) \citep{chi1997}. While implementing the AVVA approach, we found that for some variables, a coarser-grain segmenting technique might be more appropriate (for example, creativity and critical thinking as 21st-century skills), while a finer-grain technique is preferable for some variables (such as factual knowledge as knowledge and remembering as a cognitive process dimension). Since one of our goals in designing the AVVA framework is to make the steps scalable so that they can be implemented for studies involving multiple variables and can be used to find out various relations between these variables, it was pertinent to keep the grain size uniform for all the variables so that analysis of various relations among variables can be done consistently. To implement this naturally, the grain size of the verbatim has been maintained consistent with the transcription process, which involves changing speakers and inserting timestamps after each line of transcription, resulting in a 3-5 second continuity on average. Although this approach may seem redundant for some variables, we have opted for it to make the implementation of the step more consistent. To maintain consistency in the semantics of variables, a half-hour video segment can be divided into distinct episodes or activity types. For instance, a 30-minute classroom recording may be segmented into categories such as objectives, instructional Activities, pre-assessment, discussion and teaching, assessment, and assessment conversation. This structured approach enables the systematic capture of all relevant information related to the variables of interest while ensuring a high degree of accuracy and precision in the coding process.

\subsubsection{Correspondance}
There should be a correspondence between the grain size of the analysis and the research question one is asking. A more appropriate unit might be the reasoning chain, which usually involves a grain size of several sentences \citep{chi1997}. In order to ensure consistency in the treatment of variables in the reasoning chain, we suggest an encoding strategy whereby all lines of dialogue within the conversation are represented at an equivalent level of granularity. While this approach may result in an over-representation of certain variables within the dataset, the underlying research inquiry is not expected to suffer from any adverse impact. This approach greatly increases scalability and applicability. For example, if our aim is to discern potential correlations between the variables in question, keeping a consistent grain size for all possible pairs of variables to calculate correlation among them makes life much easier. Although, as suggested by \citep{chi1997}, experiments should be done with different grain sizes to ensure reliability.

\subsubsection{Features used for segmenting}
 The boundaries of analytical units can be determined using either non-content features or semantic features, similar to how subsets of protocol data are selected. Non-content features may be identified through: (a) language-related syntax, such as individual words, complete sentences, the presence of connecting terms like because or therefore, or the inclusion of equations; and (b) activity-related indicators, including pauses of a specific duration, shifts in turn-taking, or transitions between activities, for example, moving from problem-solving to consulting the text, from reading to drawing diagrams, or from reading to writing equations \citep{chi1997}. In applying the AVVA method, we suggest the use of non-content features for the purpose of segmentation. Specifically, activity-based features can be utilised, encompassing aspects such as pauses exceeding a certain duration, turn-taking dynamics, and transitions between different activities or tasks. These activities may include problem-solving, information-seeking from textual sources, transitioning between reading and diagramming, or shifting from reading to equation formulation, among others. These activity-based cues serve as crucial markers for segmenting the data. 

\subsection{Developing coding schemes}
The next step is the process of devising a coding scheme, which can be quite challenging for a researcher. Although an initial coding scheme may be constructed from the literature, published research is unlikely to provide a perfect model which can be re-applied: the choice of codes will depend upon the subject domain, the hypotheses being tested, the research questions being asked, and the ‘theoretical orientation’ of the researcher \citep{chi1997}. Content analysis is considered a flexible technique for examining textual data. In conventional content analysis (Inductive category application), categories emerge directly from the data during analysis, allowing the researcher to develop a deeper understanding of the phenomenon. In contrast, directed content analysis (deductive category application) begins with an initial coding framework based on existing theory or prior research; as the analysis progresses, new codes may be added and the scheme refined, enabling researchers to extend or clarify existing theoretical frameworks. The summative approach differs from both by focusing on the frequency or patterns of specific words or content segments, which are then interpreted to reveal their contextual meaning \citep{Hsieh2005ThreeAT}. Any of these three methods can be applied according to the needs of the study. In the implementation of the AVVA framework, we adopted directed content analysis for all the variables. A thorough literature review was done to understand various taxonomies and coding schemes already existing in the literature, and then one of the coding schemes was chosen on the basis of its comprehensiveness and ease of applicability. Generally, well-established taxonomies or coding schemes should be chosen so that lots of examples of use cases are available, which facilitates the next step, which is developing operational coding. 

\subsection{Developing operational coding schemes}
After establishing a coding scheme, the next step in verbal analysis involves determining which utterances within the verbal data serve as evidence for inclusion in a particular category or correspond to a specific code. This section provides several examples demonstrating how particular units of speech align with their representative codes. It also addresses two key challenges that may emerge during this process: resolving ambiguities in interpretation and deciding the extent of contextual information required for accurate interpretation. In the AVVA approach, we call this developing operational coding scheme. For example, in this study, generally taxonomies are available for many of the variables, hence it falls under deductive content analysis. Now, on the application level, we found that there is a need to develop an operational coding scheme based on the taxonomies given in the literature. This can be created in the presence of experts and assistants involved in the research through discussions. In using the AVVA technique for this study, it required 2-3 months to create an operational coding document for each variable. In this study, the two issues, ambiguities in interpretation and deciding the extent of contextual information required for accurate interpretation, can be addressed in the following way.

\subsubsection{Ambiguity}
The coding scheme for certain variables, such as Cognitive Process and Knowledge, follows a relatively straightforward approach primarily based on keyword identification. However, variables like Learning Theories and 21st Century Skills present challenges due to the occurrence of intricate chains of arguments, leading to inherent ambiguity in interpretation. Consider the variable Learning Theories, where a convergence of instructional strategies advocated and utilised by cognitivism and behaviourism can be observed, albeit with distinct underlying motivations. Notably, a salient commonality emerges in the utilisation of feedback. Behaviourists employ feedback (reinforcement) to shape behaviour in a desired direction, while cognitivism leverages feedback (knowledge of results) to guide and facilitate accurate mental connections \citep{Thompson1992}. In situations where inherent ambiguities arise, we have implemented a coding approach that accounts for the possibility of overlapping perspectives or intentions within the same lines of dialogue. Specifically, when encountering instances where it is unclear which specific learning theory the teacher intends to implement in the classroom, we have encoded those lines as both Behaviourism and Cognitivism. By adopting this approach, we acknowledge the uncertainty surrounding the teacher's intentions and ensure that both possibilities are captured within the coding scheme. This allows for a comprehensive analysis that recognises the potential coexistence of multiple learning theories within the same instructional context without prematurely assigning a definitive categorisation based on ambiguous or unclear indications. In the implementation of this framework to support coder comprehension and ensure consistent implementation across the dataset, the operational rules developed for all the variables were grouped into the following categories. Decision rules: They define the broad principles for encoding a variable, closely reflecting the original theoretical coding scheme and guiding consistent application during analysis. Inclusion criteria: These specify what must be present in a segment for it to be coded in a particular category. It consists of positive indicators, signals and features that confirm the category. Exclusion criteria: These specify what must not be present, i.e., conditions under which a segment should not be assigned to the category, even if it looks similar. It includes ambiguous cases that resemble the category but do not qualify, cases that superficially match but belong elsewhere, etc. Boundary conditions: These specify interpretive limits, meaning conditions where both categories may appear plausible, rules to resolve borderline cases and clarifications about contextual dependence. They can be used for: tricky cases (e.g., analogies, examples, cross-domain terms), places where coders typically disagree, conditions requiring secondary checks, etc. Hierarchical priority rules: These tell coders what to choose when multiple categories appear plausible.

\subsubsection{Context}
A secondary technical concern relates to the consideration of context during the coding process. Context, in this context (no pun intended), refers to the extent to which the coder should take into account the surrounding lines of the protocols (both preceding and subsequent) when interpreting the meaning of the current segment. The interpretation of a specific segment can vary depending on whether a broader context, encompassing several lines of transcribed verbatim, is considered or if the coding decision is confined to a very localised level, focusing entirely on the immediate segment. \citep{chi1997} proposed two primary approaches to managing this concern. The first employs minimal context, where each segment is coded based on its immediate, localised features, called strictly Local context coding. This approach proves particularly suitable for multi-participant settings, where attending to extensive surrounding discourse can introduce noise and reduce coding reliability. The second approach advocates for broader context coding, where coders consider a wider temporal window when interpreting each segment. \citep{chi1997} further recommended dual coding, which suggests encoding the entire dataset twice, once with strictly local context and once with broader context, to capture how interpretive meaning shifts across contextual scales. While this approach offers theoretical comprehensiveness, it is highly labour-intensive and may introduce additional ambiguity, particularly in complex, multi-participant classroom environments with overlapping talk and activity structures \citep{erickson2006}. 

In the implementation of the AVVA framework, we addressed the context issue through a multi-level strategy that integrates contextual information without requiring dual coding. First, we employed data triangulation through multimodal analysis: rather than relying solely on transcripts, coders accessed synchronised audio-video recordings alongside verbatim transcripts. This multimodal approach enabled coders to interpret utterances within their full communicative context, incorporating nonverbal cues (gesture, gaze, facial expression), spatial positioning, and material artefacts (e.g., diagrams on the board, manipulatives). This triangulation of verbal and nonverbal data types provided essential context for clarifying meaning without requiring a separate broader-context coding pass of the transcript alone.
Second, for variables that theoretically require sequential understanding, such as argument chains, extended problem-solving episodes, or multi-turn critical thinking sequences, etc., we encoded entire sequences as units rather than fragmenting them into isolated utterances (Chunking). This approach incorporates broader context within the variable definition itself, ensuring that codes reflect the full semantic and pragmatic meaning of extended interactions.
Third, we operationalised macro-level pedagogical context through global episode encoding. Each classroom session was segmented into three instructional phases: pre-assessment, teaching, and assessment. These global variables capture the overarching lesson structure and provide a macro-contextual frame within which micro-level codes can be interpreted. For example, cognitive processes such as understanding-conceptual knowledge pairs are more likely to occur during the teaching phase than during pre-assessment, and hence, this can be used to clarify the context. By correlating our line-level codes with these global episode variables, we can identify not just what behaviours occur, but when and in what instructional context they occur. 

This multi-level approach addresses \citeauthor{chi1997}'s (\citeyear{chi1997}) concern about contextual interpretation while remaining methodologically suitable for large-scale video datasets. By triangulating verbal and nonverbal data sources, encoding sequential variables as units, and capturing macro-context through global variables, we integrate both fine-grained and broader contextual information without the redundancy and interpretive inconsistency risks associated with dual coding. This strategy aligns with established principles of data triangulation \citep{denzin1978, patton1999} and multimodal discourse analysis \citep{kress2006, Norris2004}, which emphasise that meaning-making in classroom interaction is inherently multimodal and context-dependent. 

\subsection{Validation and reliability}
\subsubsection{Validation}
Once an operational coding system has been developed, its reliability and validity should be examined. Validity addresses the fundamental question of whether a measurement accurately reflects what it is intended to measure. According to \citep{Moskal2000-tu}, validity represents the extent to which available evidence supports the correctness of these interpretations and the appropriateness of their intended use. In the AVVA technique, we recommend that after developing an operational code, it should be checked for the accuracy of these interpretations by an expert in the subject for its content validity, the degree to which coding categories adequately represent the theoretical constructs being measured, is typically 
established through expert review \citep{haynes1995, lynn1986}. The operational coding scheme developed based on coding scheme in previous step of AVVA framework is reviewed for the coding manual, operational definitions, and example coded segments to assess whether:  (b) the interpretation of the coding scheme to create operational coding was sufficiently clear and unambiguous, and (c) the scheme comprehensively covered the range of behaviours observed in classroom video data. 

\subsubsection{Reliability}
Researchers across disciplines frequently need to assess the quality of their data collection methods. In many studies, instruments such as survey questionnaires, laboratory procedures, or classification systems are administered by multiple individuals, often referred to as raters, observers, or judges. To minimise the influence of rater variability on data quality, it is essential to determine whether all raters apply the method consistently. Inter-rater reliability measures the degree of agreement among raters when evaluating the same participants; higher similarity in their scores indicates greater reliability of the data collection process \citep{Gwet2008-xt}. Several statistical measures are available to assess inter-rater reliability. Common examples include percent agreement \citep{McHugh2012InterraterRT}, Cohen’s kappa (for two raters) \citep{Cohen1960ACO}, Fleiss’ kappa (an extension of Cohen’s kappa for three or more raters) \citep{Fleiss1971MeasuringNS}, the contingency coefficient, Pearson’s r, Spearman’s rho, the intra-class correlation coefficient, the concordance correlation coefficient, and Krippendorff’s alpha, which is particularly useful when multiple raters and rating categories are involved. However, correlation-based measures such as Pearson’s r may provide misleading estimates of agreement, potentially exaggerating or underestimating the true level of concordance among raters \citep{McHugh2012InterraterRT}. A sound methodological recommendation for researchers is to compute both per cent agreement and Cohen’s kappa when assessing interrater reliability. In situations where there is a higher likelihood of random guessing among raters, the kappa statistic provides a more appropriate correction for chance agreement. Conversely, when raters are well trained and the probability of guessing is minimal, per cent agreement alone can serve as a sufficiently reliable indicator of interrater consistency \citep{McHugh2012InterraterRT}.
\subsubsection{Training procedure for establishing Inter Rater Reliability}
Preliminary calibration phase: Each participant (student coder) is allotted approximately two hours (variable depending on the coding framework) to study the developed operational coding scheme. Following this, they are given one hour to practice coding a short transcription sample of 5–7 minutes duration. Cohen’s Kappa ($\kappa$) and percentage agreement is then calculated to assess the agreement between the participant’s coding and the reference (researcher’s) coding. If the $\kappa$ value is below 0.4 and the percentage agreement is below 80, indicating insufficient agreement, the participant proceeds to the Advanced Calibration Phase. Otherwise, they can directly participate in the main experiment.

Advanced calibration phase: If the $\kappa$ value remains below 0.4, it suggests that the participant requires additional exposure and conceptual clarity to improve their understanding of the coding framework. Accordingly, participants are given one full day to review supplementary materials, such as the reference text for the different variables. For some variables, the researcher might conduct a detailed discussion to clarify their conceptual meanings and identification during the coding process. After this additional training, Cohen’s Kappa is recalculated. If the $\kappa$ value continues to fall below 0.4, feedback is collected from participants and subject experts to identify potential sources of disagreement. Based on these insights, the coding modules are refined to enhance clarity and consistency before proceeding with the actual experimental coding phase.

\subsection{Representing the coded data}
Once the coding process is complete, the results must be depicted in a systematic way. This representation serves two primary purposes. First, it is a crucial method for presenting the data and the subsequent analysis to an audience, much like quantitative data is presented in tables or graphs. Second, and more central to the analytical process, depicting the data allows the researcher to detect patterns, trends, and structures that may not be apparent from the raw encoding alone. Given the multimodal and temporal nature of video data, several complementary methods of representation can be used, ranging from simple frequency and duration counts to complex graphical models. A straightforward and effective way to represent coded data is through tabular formats or frequency plots, such as bar charts, pie charts, etc. While applying this framework, we used raw frequency counts vs variable, overall percentage frequency vs variables and total percentage time spent on each variable. While frequency-based analysis quantifies the occurrence of instructional variables, duration-based analysis represents their temporal extent and depth, thereby providing a more nuanced and comprehensive understanding of teaching–learning processes. If the coding scheme involves a taxonomy of categories (e.g., types of gestures, categories of verbal statements, emotional expressions), a simple table presenting the frequency or mean occurrence for each category can provide a powerful, high-level summary of the dataset. For large-scale analyses, this method is particularly valuable as it can distil vast amounts of complex data into an easily digestible format, highlighting the most and least common events across the entire corpus.

However, frequency plots provide a static, aggregated view and do not leverage one of the key advantages of video data: precise temporal information. By using the timestamps captured during transcription and coding, time plots can be generated to visualise when specific coded events occur over the duration of the video. A time plot might represent the entire video duration on the x-axis and mark the occurrence of different codes on the y-axis. This representation is powerful for uncovering dynamic patterns, such as the sequence of actions, the rhythm of a conversation, or the temporal relationship between different modalities. For instance, a time plot could reveal whether a particular hand gesture consistently precedes a specific type of verbal explanation or if expressions of confusion are clustered around certain segments of a task. When used together, frequency and time plots provide a more complete picture. A frequency plot might reveal that "clarifying questions" is a high-frequency code, but it cannot tell you when those questions occurred. A corresponding time plot could show that these questions were clustered in the first two minutes of the interaction, suggesting an initial phase of orientation, or that they appeared at regular intervals, suggesting a pattern of iterative sense-making. The frequency plot answers "how much," while the time plot answers "when and in what sequence."

To depict the underlying structure of the data, it is often necessary to employ more sophisticated graphical representations. The selection of an appropriate representation depends on the research objectives and the nature of the knowledge being examined. For instance, Semantic Networks, Argument Chains, and State Transition Diagrams can serve as suitable representational approaches in different analytical contexts.

\subsection{Finding patterns and coherence in the represented data}
Once the coded data has been depicted in various formats, from simple frequency plots to complex graphical models, the next critical step is to seek out meaningful patterns and assess the coherence of the underlying structure. This process is analogous to the analysis of quantitative data, where one looks for linear trends, functional relationships, or significant differences. However, given the richness and dimensionality of audio-video data, this framework employs a multi-layered analytical approach, progressing from foundational correlational analyses to advanced computational techniques to uncover increasingly subtle and complex patterns.
\subsubsection{Statistical association}
The initial stage of analysis employed statistical techniques to examine potential relationships among the coded variables. The choice of a correct statistical test should be driven by the nature of the research questions. The examination of the association between variables depends fundamentally on the measurement level of the data and the distributional properties of the variables \citep{field2013, tabachnick2019}. To analyse the relationship among variables, researchers should choose between parametric and non-parametric approaches based on whether key statistical assumptions are satisfied \citep{siegel1988nonparametric}. 

While applying Parametric approaches, the most common assumptions regarding data drawn from a population with a specific distribution are normality and homogeneity of variance \citep{howell2012statistical}. Parametric approaches offer greater statistical power and precision if these assumptions are met. For continuous variables measured at the interval or ratio level that resemble normal distributions, Pearson's product-moment correlation ($r$) is appropriate for examining linear relationships \citep{cohen2003applied}. When data meet normality and homoscedasticity assumptions, for comparing means across groups, parametric tests such as independent samples $t$-tests, one-way ANOVA, or multiple regression are suitable \citep{maxwell2017designing}. These approaches are commonly implemented in educational research when analysing standardised test scores, interval-scaled survey responses, or other continuous outcomes \citep{huck2012reading}.

On the other hand, non-parametric approaches are suitable when the data are categorical, ordinal, or when parametric assumptions are disregarded \citep{siegel1988nonparametric, hollander2013nonparametric}. The Chi-Square Test of Independence can be used to check the statistical independence of two categorical variables at the nominal level \citep{agresti2007categorical}. For ordinal data or continuous data, Spearman's rank-order correlation ($\rho$) or Kendall's tau ($\tau$) assesses monotonic relationships without requiring linearity or normality \citep{gibbons2020nonparametric}. The Mann-Whitney $U$ test and Kruskal-Wallis $H$ test serve as non-parametric alternatives to $t$-tests and ANOVA, respectively, for comparing distributions across groups \citep{conover1999practical}. These methods are particularly important in observational classroom research, where frequency data are often non-normally distributed and measured categorically \citep{bakeman2011sequential}.

In the application of the framework for this study, all variables, namely knowledge dimension, cognitive process dimension, learning theories, and 21st century skills, are categorical variables. Furthermore, preliminary inspection of the data revealed non-normal distributions typical of event-coded classroom 
interaction data, where many episodes contained zero instances of 
specific variables, resulting in highly skewed frequency distributions. Henceforth, a non-parametric approach, the Chi-Square Test of Independence used for hypothesis testing \citep{agresti2007categorical}. This test is well-suited for analysing contingency tables constructed from binary-coded classroom variables and has been widely used in educational discourse analysis \citep{hennessy2016, mercer2010}.  The significance of pairwise relationships was evaluated through p-values. These values determined whether observed associations, for instance, between a specific 21st century skill and a learning theory, reflected genuine effects rather than random coincidence. 
% The Phi coefficient ($\phi$) was used to quantify the magnitude of association between binary or categorical variables, complementing the inferential significance indicated by p-values. 

In addition, each line of the transcription is timestamped too. Since the timestamped data is ordinal data, rather than continous and normally distributed, the Spearman rank correlation coefficient ($r_s$) was employed to assess monotonic relationships. Spearman's correlation is robust to outliers and non-normality, and unlike Pearson's $r$, does not assume 
linearity, making it particularly suitable for timestamped sequences where 
variables may exhibit non-linear patterns across episodes \citep{pett2016nonparametric}. Together, the Chi-Square Test of Independence and Spearman's rank-order correlation provided complementary perspectives: the former examined structural associations between categorical variables (whether variables co-occur more or less frequently than expected by chance), while the latter captured duration-based co-variation patterns across the episodes \citep{bakeman2011sequential}. This dual analytical approach aligns with established methodologies in classroom interaction research, where both categorical and temporal dimensions of discourse are theoretically meaningful \citep{derry2010}. As recommended by the \citep{APA2020}, the effect size should always be reported with p-values. Hence, the Phi coefficient ($\phi$) was used to quantify the magnitude of association between binary or categorical variables, complementing the inferential significance indicated by p-values. In the duration-based analysis, descriptive statistics were tabulated, showing the maximum, mean, and minimum time spent on each variable. Also, the variables' normality was assessed using the Shapiro–Wilk test. This combination of inferential testing and effect-size estimation provided both statistical rigour and interpretive depth in identifying meaningful correlations among the study’s coded variables. This analysis provides a first-pass, quantitative map of the most direct relationships within the dataset. 

\subsubsection{Interaction metrics}
 Building on this, the analysis can move to more complex interaction parameters designed to test specific hypotheses derived from theory. For example, to understand the relationship between two variables, one might define a custom interaction parameter like an EPSI index \citep{Upadhyay2025}. Such an index could be formulated to quantify the interaction between variables, learning theories and 21st century skills. By analysing how this index varies across different episodes, we can move beyond simple correlations to investigate more nuanced, theory-driven questions about how specific pedagogical frameworks affect the occurrence of 21st century skills.

\subsubsection{Multivariate pattern analysis and unsupervised learning: General approaches}

When coded classroom discourse data exhibit complex, high-dimensional structure, multivariate statistical techniques and unsupervised machine learning methods can reveal latent patterns that univariate analyses may miss \citep{baker2014big, siemens2013learning}. Multivariate pattern analysis treats each transcript segment as a feature vector, comprising frequencies of variables, transition probabilities between codes, temporal sequences, and other quantifiable attributes, and applies dimensionality reduction or clustering algorithms to identify underlying structure \citep{bogarinromero2018clustering, ochoa2021multimodal}.

\paragraph{Principal component analysis (PCA)}: It reduces high-dimensional data to a smaller set of orthogonal components that capture maximum variance, enabling visualisation of complex patterns in two or three dimensions \citep{jolliffe2016principal}. In educational contexts, PCA has been used to identify latent factors in student behaviour \citep{bowers2017teacher} and to reveal groupings of instructional strategies \citep{kelly2017learning}. In this study, since each of the four variables have sub variables, for example cognitive process dimension has six sub-variables ('remembering', 'understanding', 'applying', 'analysing', 'evaluating', 'creating'), and 21st-century skills have 4 sub-variables ('problem solving', 'creativity', 'collaborative learning', 'critical thinking) etc., summing to 17 main variables and then over arching variables like instructional phases (global encoding), grades (6th-12th), and topics (science and mathematics) brings 12 more variables which makes this dataset high dimensional. Hence, PCA was implemented to understand the complex patterns among these variables by reducing dimensionality. 

\begin{figure}[h] % 'h' means place it here
    \centering
    \includegraphics[width=0.9\textwidth]{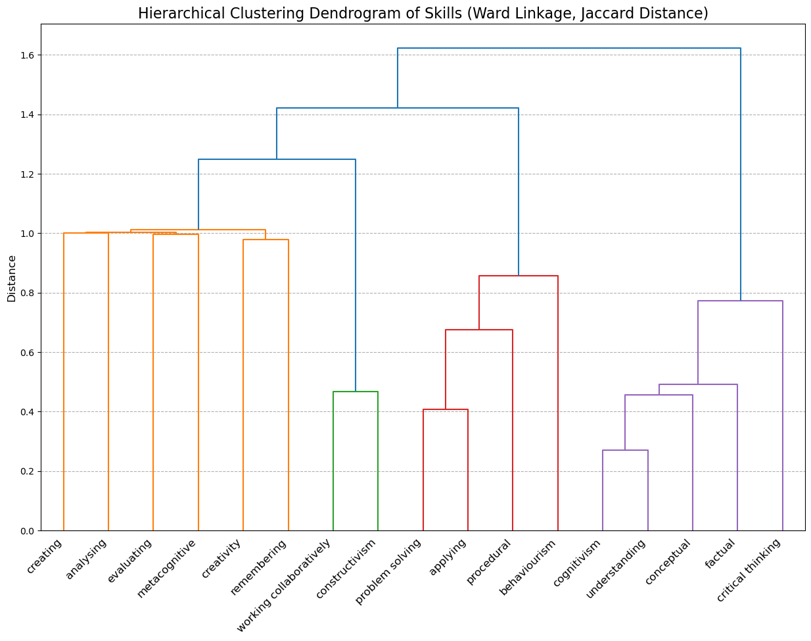} % Replace with your image file
    \caption{Example of the Hierarchical clustering Dendogram for all the variables using Ward linkage and Jaccard distance}
    \label{fig:2}
\end{figure}

\paragraph{Clustering algorithms}: "Data clustering is a widely used data processing technique that groups items into distinct clusters based on some measure of similarity between data points" \citep{Ran2022, Singh2024}. The three important uses of data clustering are as follows: Highlighting the patterns and astuteness in the data, recognising the degree of similarity between data points and data organisation and summarisation through cluster prototypes \citep{Singh2024, Celebi2013}. Clustering algorithms are broadly classified into two categories: Hierarchical Clustering Algorithms and Partitional Clustering Algorithms \citep{Singh2024, Ezugwu2022}. Partitional clustering algorithms are further divided into several subdivisions, including density-based (e.g., DBSCAN), model-based (e.g., GMM), and square-error-based (e.g., K-means). Hierarchical clustering builds nested dendrograms revealing multi-level structure \citep{murtagh2012algorithms}. Density-based clustering methods such as Density Based Spatial Clustering of Applications with Noise (DBSCAN) are traditional and popular clustering methods used to extract hidden patterns from datasets by separating high and low density regions based on the neighbourhood information \citep{Maheshwari2023}. Gaussian mixture models (GMM) partition data into homogeneous groups based on similarity metrics \citep{hastie2009elements}. K-Means clustering is one of the simplest and widely used unsupervised machine learning algorithms, which is a centroid-oriented partition scheme that groups data points into a predetermined number of clusters from the given data points \citep{macqueen1967methods, Singh2024, ShiYu2012}. Clustering algorithms have been applied across a range of educational research contexts: agglomerative hierarchical clustering has been used to check if student groups can be detected based on the students' distinct use of learning strategies in a flipped classroom
\citep{Jovanovic2017}; $K$-means clustering has been applied to
group students by behavioural patterns recorded in the learning management system event logs \citep{Krizanic2020}; Cluster analysis techniques like k-means ++, fuzzy k-means, DBSCAN were applied with the aim to find out the groupings of students at three points in the teaching-learning process (initial, intermediate, and final), in Moodle-based course monitoring \citep{Baz2021}; and Gaussian Mixture Models have been used to capture uncertainty and provide probabilistic groupings of student engagement levels in virtual learning environments, outperforming the $K$-means clustering method in cluster analysis \citep{Nimy2023}. In this study, we applied various clustering techniques to understand how the occurrence of variables is grouped in the classroom discourse analysis. Figure~\ref{fig:2} shows a sample implementation of the hierarchical technique for the dataset.\\ 

Some other techniques, such as Latent Semantic Analysis (LSA) and Topic modelling (e.g., Latent Dirichlet Allocation (LDA)), extract thematic structure from text by identifying co-occurrence patterns of words or codes \citep{blei2003latent, landauer1998introduction}. These techniques have been successfully applied to classroom transcripts to identify dominant discourse themes \citep{dowell2018modeling} and predict learning outcomes from discussion content \citep{chen2019automated}.
In the analysis of student evaluation comments, LDA has been employed to extract topics from open-ended feedback, revealing patterns in student perceptions of teaching and learning experiences without the labour-intensive manual coding process \citep{yang2023using}.

\subsubsection{Machine learning approaches for educational text analysis}
Various Machine Learning (ML) approaches have been used for text analysis in the educational context. Support Vector Machine (SVM) have demonstrated strong performance in various educational text classification tasks due to its effectiveness in handling high-dimensional feature spaces. They have been employed to classify online education resources, where an SVM-based classifier achieves slight improvements in both precision and recall ratios for resource classification, compared with traditional neural network and deep learning baselines \citep{osmanoğlu2022improved}. In sentiment analysis of student feedback, ML-based models have been included in comparative evaluations, including classifiers such as SVM, Random Forest (RF), Stochastic Gradient Descent (SGD), Multilayer Perceptron (MLP), and Multinomial Naive Bayes (MNB), and Lexicon-based models often demonstrate robust performance across diverse educational datasets \citep{acheampong2024sentiment,deshpande2025elevating}. For argumentation mining in educational discourse, as one of the methods, SVMs have been used to identify argument components in student essays, leveraging their capacity to handle complex feature representations derived from argumentative structures \citep{habernal2017argumentation}. ML models, such as RF, SVM, Naive Bayes (NB) and Decision Tree (DT), have been combined with feature extraction techniques such as Recurrent Neural Network (RNN) and Long Short-Term Memory (LSTM), to predict the performance of students as pass or fail \citep{kukkar2024novel}. In the context of student comment analysis, RFs have been used alongside other ensemble methods to classify feedback sentiment into positive, negative, and neutral categories, demonstrating strong performance when combined with Term Frequency-Inverse Document Frequency (TF-IDF) feature representations \citep{nasim2017sentiment,acheampong2024sentiment}. The gradient boosting techniques have also been applied to improve classification accuracy in educational text analysis. When combined with hybrid feature representations that integrate RoBERTa embeddings with handcrafted linguistic features, LightGBM (a variant of gradient boosting) achieved an enhanced scoring precision in automated essay scoring tasks \citep{hybrid2024roberta}. These methods demonstrate particular strength in handling high-dimensional feature spaces and modelling complex feature interactions in educational text data.

\subsubsection{Deep learning approaches for educational text analysis}
Advancements in deep learning have opened new opportunities for modelling the complex patterns in educational datasets \citep{torkhani2024oulad}. Convolutional Neural Networks (CNNs), though originally designed for image processing, have demonstrated significant success in text classification by identifying local n-gram patterns. These models employ 1-D or 2-D convolving filters to slide across word embedding matrices, capturing essential key phrases and syntactic structures \citep{Soni2022}. In automated essay scoring, CNNs have been employed to extract hierarchical features from essay text, often in combination with recurrent layers to capture both local patterns and long-range dependencies \citep{taghipour2021review,dasgupta2018augmenting, kusuma2020cnn}. RNNs are specifically designed to exploit the sequential characteristics of Language. By maintaining a recurrent cell that processes input one token at a time while retaining a hidden state, RNNs create a form of "memory" that informs decisions based on prior context \citep{JurafskyMartin2024SLP3}. LSTM networks have emerged as particularly effective architectures for modelling sequential patterns in educational text. In automated essay scoring, LSTMs have been widely adopted, with studies demonstrating superior performance compared to traditional recurrent networks in predicting student performance from sequential assessment data \citep{torkhani2024oulad}. BiLSTMs (Bidirectional LSTMs) have been integrated with attention mechanisms to prioritise salient components of essays, with SBERT embeddings feeding into BiLSTM layers for enhanced semantic understanding \citep{sbert2024automated}. Gated Recurrent Units (GRU) offer a computationally efficient alternative to LSTMs for sequential text modelling in educational applications. One study employed a GRU architecture with dense layers, max-pooling, and ADAM optimisation to predict students in need of academic assistance, achieving superior accuracy compared to traditional RNN and AdaBoost models \citep{advanced2024gru}.\\

Pre-trained word embeddings have become fundamental components of modern educational text analysis systems. Word2Vec embeddings, using either CBOW or Skip-gram architectures, have been employed to capture semantic relationships in student writing, enabling a more nuanced understanding of essay content beyond bag-of-words representations \citep{Mikolov2013EfficientEO}. GloVe embeddings, a log-bilinear regression model for the unsupervised learning of word representations, leverage global word co-occurrence statistics, which outperforms other models on word analogy, word similarity, and named entity recognition tasks \citep{pennington2014glove}. FastText embeddings handle out-of-vocabulary words through character n-gram representations \citep{Bojanowski2017}. These embedding techniques serve as foundational layers in many deep learning architectures for educational text analysis.\\

In classroom discourse analysis, BERT-based classifiers have been employed to compare the use of Transformer-based language models for the automatic classification of teaching behaviours from real classroom transcriptions \citep{MartnHoz2025}. Recognising the importance of capturing intricate student–teacher interactions, a study reported that Sentence-BERT (S-BERT) bi-encoder architectures significantly outperform traditional cross-encoder models such as BERT and RoBERTa in classifying high-inference discourse elements, including High Uptake and Focusing Questions \citep{lee2025investigating}. For automated essay scoring, BERT encodings have been integrated with LSTM-based spatiotemporal processing to combine semantic text interpretation with behavioural features \citep{deepgrade2025iot}. DistilBERT, a distilled version of BERT, has demonstrated the highest performance in sentiment analysis of student feedback, when applied through transfer learning \citep{distilbert2024sentiment}. Studies have employed BERT-based models to automatically detect student talk moves (dialogic acts) in mathematics lessons, comparing their performance with ChatGPT and finding that fine-tuned BERT models outperform large language models in accuracy while lacking the interpretability that LLMs provide through explanations \citep{chen2023chatgpt}. Research comparing BERT and Llama3 for dialogic analysis in teacher professional development found that BERT exhibited substantially higher accuracy in analysing dialogic moves, though both models contributed to teachers' learning of dialogic pedagogy \citep{wang2025evaluating}. 

\subsubsection{Computational efficiency and scalability}
The time investment required for manual coding is substantial. In our study, developing the operational coding scheme through iterative expert review required approximately 2–3 months. Training coders and establishing inter-rater reliability added another 3–4 weeks. Encoding 48 classroom episodes (approximately 24 hours of video) required approximately 1 month of dedicated coding work by multiple trained coders. The total timeline from coding scheme development to a fully encoded dataset spanned approximately 4–5 months for one variable.\\
While our study employed manual coding, we recognise the potential of NLP-assisted approaches for future large-scale analyses once initial training is complete. The base model used in this study is the 4-bit quantised \texttt{unsloth/llama-3-8b-Instruct-bnb-4bit} variant of LLaMA 3 \citep{unslothllama32024}, fine-tuned using supervised fine-tuning (SFT) with the TRL library \citep{vonwerra2020trl} and optimised via the Unsloth framework \citep{unsloth} built on Transformers. Preliminary experiments (not reported here) conducted on the encoded dataset indicate that fine-tuning pre-trained transformer models (e.g., BERT-base and LLaMA-3) on 20,000–30,000 labelled utterances, using a consumer-grade GPU (NVIDIA RTX 4060 with 8 GB VRAM), required approximately 12–24 hours per variable (e.g., Critical Thinking). This setup achieved reasonably good performance, with a precision of 0.75, a recall of 0.76, and an F1-score of 0.71 with the LLaMA 3 model for the variable critical thinking. Once the models are trained, encoding new transcripts is nearly instantaneous; a trained model can classify thousands of utterances per minute. For researchers analysing large video corpora (e.g., hundreds of hours), this efficiency gain is transformative. However, deep learning methods require substantial upfront investment in creating labelled training data. A hybrid approach, manually coding a representative subset of transcripts to train an automated classifier, then using the classifier to code the remainder with human verification, can balance accuracy and efficiency.

\subsection{Interpreting the pattern and its validity}
Following \citeauthor{chi1997}'s (\citeyear{chi1997}) guidance for interpretation, pattern discovery was always linked back to our theoretical framework and research questions. In practice, we triangulated the findings across multiple perspectives: theoretical conjectures, statistical confirmation, and machine-learning outcomes, etc.

\subsubsection{Triangulation of perspectives}
In this study, after identifying patterns in the data, we evaluated whether they aligned with established cognitive and learning theories, etc. This step served two purposes: 
First, it helped confirm that observed correlations reflected genuine educational 
phenomena rather than statistical artefacts; second, it provided theoretical 
grounding for interpreting what these patterns mean for classroom practice. For each significant correlation, we asked: Does this relationship make sense given what the literature says? Patterns that aligned with prior research were interpreted as robust findings, while correlations lacking theoretical support were treated cautiously and examined more closely. For example, we found strong correlations between Applying vs Procedural Knowledge, and between Understanding vs Conceptual Knowledge. These relationships are consistent with Anderson and Krathwohl's revised Bloom's taxonomy \citep{Anderson2000-jt}. Similarly, our finding that constructivist teaching practices correlated with collaborative learning mirrors patterns reported in the education literature \citep{HmeloSilver2004}. These theoretically grounded relationships received greater weight in our interpretation. In contrast, when we identified correlations without a clear theoretical justification, we investigated further to determine whether they might reflect confounding variables, instructional phase-specific effects, or statistical chance. This approach, using theory as a lens to interpret statistical patterns, ensured that our conclusions were not only empirically supported but also educationally meaningful.

\subsubsection{\raggedright Statistical validation (grain-size experiments, bootstrapping and stability assessment)}

As suggested by \citep{chi1997}, a grain-size test should be implemented by repeating the full encoding process at an alternative grain, treating convergence of qualitative patterns as evidence of validity. In the present framework, we operationalise this recommendation computationally and extend it so that the approach scales to large corpora without adding to the burden of the primary coding task, which is itself among the most time-intensive activities in observational research. As described in the previous steps 3 and 4, the grain size of the verbatim is taken as the same as that generated during the transcription process, which was based on activity features, and chosen as the natural base unit of analysis. All grain size experiments are then post-hoc aggregations of this natural base unit. While doing the grain size experiments, we realised that correlation strength often varied systematically with the size of the time unit discussed in the next section. 
\paragraph{Modifiable temporal unit problem}: A methodological challenge in analysing time-stamped sequential data involves the choice of the "grain size", a temporal aggregation window, at which observations are partitioned for analysis \citep{openshaw1984modifiable, cheng2014modifiable}. This challenge, known as the Modifiable Temporal Unit Problem (MTUP), is the temporal analogue of the well-documented Modifiable Areal Unit Problem (MAUP) in spatial statistics \citep{fotheringham1991modifiable}. The MTUP describes how statistical relationships between variables can sometimes dramatically vary depending on the temporal resolution at which data are aggregated \citep{cheng2014modifiable, swift2008choice}. MTUP can have three types of effects.  
\paragraph{Temporal aggregation effect}: It is a process that involves the discretisation of the time frame from a detailed interval into a coarse one \citep{cheng2014modifiable}.
In discourse analysis, this problem exhibits when deciding whether to aggregate coded variables at the utterance level, the turn level, fixed time windows (e.g., 30s group), or activity-based episodes \citep{chi1997, bakeman2011sequential}. Correlation coefficients, regression parameters, and other statistical summaries can be highly sensitive to this choice \citep{dark1997inference}. In extreme cases, a correlation that appears positive at one temporal scale may become negative at another, a phenomenon known as Simpson's Paradox \citep{simpson1951interpretation, wagner1982simpson}, which represents a specific manifestation of the broader Ecological Fallacy where relationships observed in aggregated data differ from those in disaggregated data \citep{robinson1950ecological, piantadosi1988ecological}. For example, in this study, say, that Cognitivism and Understanding co-occur, the strength of that association necessarily depends on the temporal window within which
“co-occurrence” is operationalised. At a fine grain (e.g., 10s), two variables will appear anti-correlated; whereas at a coarse grain (e.g., 180s), they will appear strongly positively correlated, yet the underlying pedagogical reality is unchanged.
\paragraph{Temporal segmentation effect or the zoning effect}: In purely temporal analysis, this effect is concerned with the starting point of the interval, and the segmentations could be different if the starting points of the intervals are different \citep{cheng2014modifiable}. In this study, we are using non-overlapping chunks starting from the first line of our dataset. However, if we were to repeat your analysis for n=10 but start the first chunk on line 2, then line 3, and so on, the zoning effect would likely produce slightly different Phi coefficients for each starting point. While our current methodology holds the zoning constant, acknowledging its existence is important for a complete methodological discussion.
\paragraph{Temporal boundary effect}: In temporal analysis, this effect is concerned with the starting and end points of a time series as its temporal extent or boundary. Changing the temporal length will produce different descriptive statistics \citep{cheng2014modifiable}. In this study, it is kept constant.

\paragraph{Implementation of temporal aggregation effect}: In this study, line-level codes were aggregated into non-overlapping time windows of width $\Delta$ seconds, separately within each instructional phase (Preassessment, Teaching, Assessment) and each episode. Episode-relative time was obtained by subtracting the timestamp (wall-clock time as shown in Figure~\ref{fig:1} ) of the first line from the timestamps of all subsequent lines within the episode. To assign a window index to a line, the episode-relative time for each line is divided by $\Delta$, followed by discretisation using the floor function. Within each window, a variable was coded as present (1) if it appeared on at least one constituent line, and absent (0) otherwise, according to the standard interval-recording convention \citep{bakeman2011sequential} also called as Logical-OR aggregation. Grains of $\Delta \in \{5, 10, 15, 30, 60, 120\}$\,s were tested.
\paragraph{Association measure}: For two binary variables $U$ and $V$ observed over $n$ windows, let
$a = |\{U{=}0,V{=}0\}|$,\;$b = |\{U{=}0,V{=}1\}|$,\;
$c = |\{U{=}1,V{=}0\}|$,\;$d = |\{U{=}1,V{=}1\}|$

\begin{equation}
  \phi = \frac{ad - bc}{\sqrt{(a+b)(c+d)(a+c)(b+d)}}
  \label{eq:phi}
\end{equation}

$\phi$ is the Pearson correlation coefficient (or phi coefficient) applied to two binary
variables and equals $\sqrt{\chi^2/n}$ in sign-adjusted form
\citep{agresti2007categorical}. It takes values in $[-1,1]$, with $\phi=0$ indicating
independence, $\phi=1$ perfect co-occurrence, and $\phi=-1$ perfect
mutual exclusion.
\paragraph{Phi ceiling and base-rate filtering}: While calculating the phi coefficient to examine co-occurrence among two variables, the maximum association that can be achieved is not always 1. It depends on the frequency with which the variables occur. If one variable occurs in nearly every observation window while the other occurs in only some windows, the phi coefficient is mathematically prevented from reaching high values even if the two variables co-occur perfectly whenever the rare one appears. This constraint was first identified by \citep{Ferguson1941} and later formalised by \citep{Davenport1991} in the educational measurement literature. In practical terms, it means that a phi value of, say, 0.06 computed for a pair where one variable fires in only 0.3 per cent of windows is not comparable to a phi value of 0.08 computed for a pair where both variables fire in 60 per cent of windows. The first is near its mathematical ceiling, while the second is not. Treating them as equivalent would be misleading.\\
To prevent this, a base-rate filter was applied before any phi computation. It is defined as the proportion of time windows in which each variable of a pair was active in the given analysis condition, i.e. within a given combination of instructional phase, observational grain size, etc. If either variable fell below a minimum threshold, the pair was excluded from phi computation in that condition. The rationale for these specific thresholds follows \citep{Bakeman1997}, who note that rare variable codes produce unreliable co-occurrence statistics, and is grounded in \citep{Davenport1991} demonstration that phi values computed under asymmetric base rates are not directly interpretable.\\
As explained above, the base-rate filter was applied to protect against the mathematical phi ceiling constraint. However, this filter should not be interpreted as a claim that the excluded variables are theoretically uninformative. For example, a correlational study at the natural base unit of analysis showed a high phi correlational value, a pattern consistent with prior findings reported by \citep{HmeloSilver2004}. Although the frequencies of occurrence for both variables are very low, their near-zero base rates in the present sample reflect the pedagogical culture of the observed Indian school classrooms rather than an absence of theoretical association. \citep{Stigler1999} and \citep{Clarke2006} document substantial cross-national variation in the prevalence of constructivist teaching methods, and the present sample should be understood as drawn from a cultural context in which behaviourist and cognitivist instruction predominates. The base-rate filter is therefore a methodological protection against statistical artefact in this specific dataset, not a theoretical claim about the relationships between the excluded variables in contexts where they occur more frequently. Researchers working in contexts where constructivism and collaborative learning are more prevalent would not face this constraint, and the phi relationships among these variables remain theoretically important even if they cannot be reliably estimated from the present data.

\paragraph{Bootstrapping: Rationale and procedure}: The standard approach to computing a confidence interval around a phi coefficient assumes that each observation is statistically independent of every other. In classroom observation data, this assumption is violated at two levels. First, observational lines recorded within the same episode share the same teacher, the same students, the same lesson topic, and the same instructional context, so they cannot be treated as independent draws. Second, within an episode, the lines are ordered in time, meaning successive observations are correlated with one another. Simply computing a standard error as if all 19,141 lines were independent would produce confidence intervals that are far too narrow and significance claims that are far too optimistic. This problem is well recognised in the observational research literature by \citep{Bakeman1997}, who explicitly note that the episode, not the individual-coded line, is the appropriate unit of statistical independence in classroom observation studies. Henceforth, we suggest episode-level bootstrapping, which is the standard resampling technique for clustered observational data \citep{Efron1993, DiCiccio1996}. Rather than resampling individual lines or individual time windows, the procedure resamples whole episodes. The bootstrap procedure was implemented as follows:

\begin{itemize}

    \item \textbf{Pre-aggregation.} Before any resampling begins, every
    episode is aggregated into fixed time windows at each grain size and
    stored in memory. A window is coded using Logical-OR aggregation. This step is
    performed once per grain size and cached, so that the bootstrap
    loop never re-reads the raw data or re-runs any grouping
    operations. This design choice reduced computation time by
    approximately two orders of magnitude compared to re-aggregating
    within each replicate.

    \item \textbf{Episode resampling}: The sampling is done with replacement
    For each of the $B = 1000$ bootstrap replicates. Because 
    episodes are the natural unit of statistical independence in classroom observation data \citep{Bakeman1997}, resampling is performed at the episode level rather than at the
    level of individual lines or windows. Some episodes may therefore
    appear multiple times in a given replicate, while others may be
    absent entirely.

    \item \textbf{Window stacking}: The pre-aggregated window arrays
    for the selected episodes are retrieved from the cache and
    concatenated into a single combined dataset for that replicate.
    No further aggregation or grouping is required.

    \item \textbf{Phi computation}: For a given pair of variable 
    The phi coefficient is computed on the combined window dataset.
    If the resulting $2 \times 2$ contingency table is degenerate,
    meaning one or both variables show no variation in the resampled
    data, so that an entire row or column of the table is zero, the
    replicate is discarded and does not contribute to the bootstrap
    distribution.

    \item \textbf{Distribution summary}: Steps 2 to 4 are repeated 1,000
    times, yielding a bootstrap distribution of phi values. The mean of
    this distribution is taken as the point estimate, and the 2.5th and
    97.5th percentiles as the lower and upper bounds of the 95 per cent
    confidence interval (CI), following the percentile bootstrap method
    \citep{Efron1993}. This method makes no distributional assumptions
    about the shape of the phi sampling distribution, which is
    important because phi distributions can be skewed when base rates
    are asymmetric \citep{Davenport1991}.

    \item \textbf{Minimum replicate threshold}: Since, degenerate tables are discarded in
    step 4, If fewer than 100 valid replicates out of 1,000 are obtained for a given variable
    pair the pair is excluded from inference entirely. This threshold
    ensures that confidence interval estimates are based 
    on a sufficient number of bootstrap draws to be
    numerically stable.

\end{itemize}

\paragraph{Stability assessment across temporal grain sizes}

When associations are measured across multiple temporal grain sizes (e.g., 10-second, 30-second, 60-second windows), evaluation of stability patterns reveals whether measured relationships are due to the choice of observational timescale or reflect grain-size dependency. Following the logic of the Modifiable Temporal Unit Problem \citep{cheng2014modifiable}, we assess stability across four distinct criteria, each capturing a different aspect of temporal robustness.

\paragraph*{Criterion 1: Sign consistency}

The sign (positive or negative) of the association should remain constant across all 
temporal grain sizes examined. Sign reversal, where an association is positive at one 
grain size and negative at another, shows timescale sensitivity. This phenomenon parallels Simpson's paradox \citep{simpson1951interpretation, Pearl2014}, in which associations can "reverse, disappear, or emerge" when data are aggregated differently. Following \citep{robinson1950ecological}, who demonstrated that correlations measured at individual versus ecological (aggregate) levels can differ dramatically in both magnitude and direction, we treat directional inconsistency across temporal grain sizes as evidence that the measured association is not a stable property of the underlying process, but rather a scale-dependent aggregation artefact \citep{cheng2014modifiable}. Sign reversal indicates that the relationship between variables is conditioned on temporal grain size rather than representing a grain size invariant relation. If a variable pair changes from positive to negative, or negative to positive, as the grain size changes, then the pattern should be examined separately at each grain size instead of being summarised across all scales

\paragraph*{Criterion 2: Confidence interval overlap}

Successive bootstrap confidence intervals (computed at adjacent temporal grain sizes) 
must share a common interval. If the 95\% bootstrap confidence interval at grain size 
$t_i$ does not overlap with the confidence interval at grain size $t_{i+1}$, the association estimates at those two scales are statistically discernible, demonstrating that temporal aggregation changes the measured relationship magnitude. Overlap should be there for every step along the temporal grain-size sequence. The use of confidence interval overlap as a consistency criterion follows \citep{Cumming2009}, who established that non-overlapping 95\% confidence intervals correspond approximately to a statistically significant difference between estimates at $p < 0.01$. Non-overlapping intervals provide direct empirical evidence that the MTUP is present, i.e. different grain sizes yield not merely sampling variations around a common true value, but substantively different association strengths. Variable pairs with non-overlapping successive confidence intervals are classified as exhibiting strong grain-size dependency, indicating that the association magnitude is a function of the temporal grain size rather than a stable phenomenon.

\paragraph*{Criterion 3: Zero exclusion}

For a variable pair to demonstrate consistent association across temporal scales, every confidence interval across all grain sizes should exclude zero (i.e., both the lower and upper bounds of each interval must be either both positive or both negative). Bootstrap confidence intervals are constructed from the empirical distribution of resampled estimates \citep{Efron1993}, and their interpretation follows standard statistical inference, where exclusion of zero implies rejection of the null hypothesis of no association \citep{CasellaBerger2002}. Extending this principle, we require consistent zero exclusion across temporal grain sizes as a robustness criterion. A confidence interval that includes zero at any grain size means that, at that time scale, we cannot say there is a relationship between the variables. Thus, if zero lies within the interval, the null hypothesis of no association cannot be rejected. In simple terms, the data at that scale are consistent with there being no real connection. If a variable pair shows this “zero crossing” at some temporal grain sizes but not others, it means the relationship is present at some temporal grain size and absent at others. This should not be seen as an error or noise. Instead, it shows that the strength of the relationship changes depending on how the data are aggregated over time. Importantly, the presence of zero at some scales does not invalidate the relationship found at other scales. It simply means the relationship is not uniform across all time scales.

\paragraph*{Criterion 4: Magnitude stability}

Even when the sign is consistent, confidence intervals overlap, and zero is excluded at all grain sizes, the interpretation of an association may change if the effect magnitude itself varies drastically across temporal grain sizes. An association of $r = 0.05$ at a fine grain and $r = 0.45$ at a coarse grain represent qualitatively different strengths of the relationship, spanning from small to large effects. Reporting a single summary value for such pairs would be misleading. We suggest magnitude stability in terms of a maximum permissible range $\epsilon$ between the highest and lowest association estimates across all temporal grain sizes. For correlation-based measures in behavioural and psychological research, \citep{Gignac2016} provide empirical benchmarks from 708 meta-analytically derived correlations: the 25th percentile (small effect) corresponded to $r = 0.11$, the 50th percentile (medium effect) to $r = 0.19$, and the 75th percentile (large effect) to $r = 0.29$. The interquartile range thus spans $0.29 - 0.11 = 0.18$. Based on these empirical distributions, we adopt $\epsilon = 0.20$ as the magnitude stability threshold for the present study. Variable pairs with $Range > \epsilon$ are called as showing magnitude-varying associations and require specific grain size reporting rather than summary statistics.

\paragraph{Pattern categories}

Rather than using these criteria as binary categories (stable vs. unstable), we group
variable pairs into different patterns based on their stability profile across the four criteria. 
This suggests $2^4 = 16$ possible patterns, each with distinct theoretical implications. Some of them are:

\begin{itemize}
\item \textbf{Grain-invariant associations}: Sign consistent, CIs overlap, zero excluded at all grains, magnitude stable. This means the relationship is consistent and does not depend on how the data are grouped over time, so it can be reasonably combined into a single overall summary.

\item \textbf{Fine-grain specific associations}: Sign consistent, CIs 
overlap, but zero-crossing occurs at coarse grains. This means the relationship is present at shorter time scales but fades when the data are grouped over longer periods, suggesting the connection mainly operates over short time intervals.

\item \textbf{Multi-scale patterns}: Sign reversal, non-overlapping 
CIs, mixed zero-exclusion, large magnitude variation. This means the relationship behaves differently at different temporal grain sizes; for example, the variables may move together at one grain size but in opposite directions at another, showing that the pattern depends on how the data are grouped over time.

\item \textbf{Magnitude-varying associations}: Consistent sign and 
statistical significance across grains, but effect magnitude changes substantially. 
These require grain-specific effect size reporting and theoretical interpretation 
of why association strength depends on aggregation level.
\end{itemize}

In the same way, different patterns can appear and give more detail about how the relationship changes over time. Each type of pattern should be understood on its own, rather than being ignored as unstable. This approach treats changes across different temporal grain sizes not as a problem, but as useful information that helps us understand how the association between variable pairs works over time \citep{cheng2014modifiable, Pearl2014}.\\
\begin{figure}[h!]
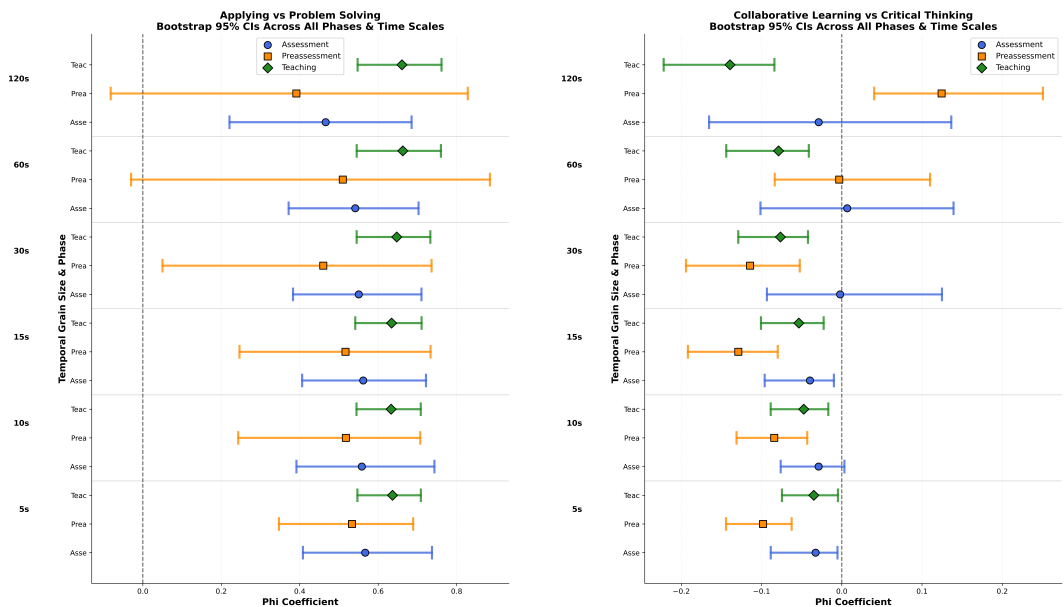

\centering
\begin{subfigure}[b]{0.48\textwidth}
    \centering
    \includegraphics[width=\textwidth]{forest_combined_Applying_vs_Problem_Solving.png}
    \caption{Sample example of stable profile for a variable pair}
    \label{fig:fig1}
\end{subfigure}
\hfill
\begin{subfigure}[b]{0.48\textwidth}
    \centering
    \includegraphics[width=\textwidth]{forest_combined_Collaborative_Learning_vs_Critical_Thinking.png}
    \caption{Sample example of unstable profile for a variable pair}
    \label{fig:fig2}
\end{subfigure}
\caption{Comparison of stable versus unstable profiles using forest plots}
\label{fig:combined}
\end{figure}
Figure~\ref{fig:combined} shows contrasting temporal stability profiles in variable associations. In Figure~\ref{fig:fig1}, Applying vs Problem Solving bootstrapping results with 95\% CIs across all instructional phases (Preassessment, Teaching, and Assessment) and temporal grain sizes (5s, 10s, 15s, 30s, 60s, and 120s) on the Y axis and the phi coefficient (size effect) on the X axis have been plotted. Since, for all three phases, Sign consistent, CIs overlap, zero excluded at all grains, and magnitude stable criteria got satisfied, this association can be classified as a grain-invariant association for the tested temporal grain sizes. This suggests that the relationship is consistent and does not depend on the grouping of data over temporal grain sizes. Similarly, in Figure~\ref{fig:fig2}, Collaborative learning vs Critical Thinking bootstrapping results with 95\% CIs across all instructional phases and temporal grain sizes have been plotted. For the preassessment phase, the phi coefficient ($\phi$) (association strength) mean value is negative at fine grains (5-30s: $\phi$ $\approx$ -0.11), near zero at 60s, and positive at coarse grains (120s: $\phi$ $\approx$ +0.14). This pattern suggests that the variables anti-correlate within brief instructional exchanges but co-occur within extended lesson phases. The magnitude range (0.27) exceeds the stability threshold ($\epsilon$ = 0.20), and the CIs at 5s and 120s do not overlap, confirming that temporal aggregation reverses the direction of the measured relationship. In contrast, the Teaching phase shows a stable negative association across all temporal grain sizes ($\phi$ $\approx$ -0.08, range $\approx$ 0.02), demonstrating phase-specific manifestation of MTUP. 

\section{Limitations and future work}
Although this study utilised audio-video recordings as the primary data source, the analytical focus remained on the verbal modality. The Audio Video Verbal Analysis (AVVA) framework was designed to extract and analyse spoken interactions through transcription and coding, integrating only minimal non-verbal information (e.g., gestures, pauses, or emphatic tones) to preserve interpretive clarity and reduce the procedural complexity of transcription, while subsequent stages of the AVVA framework are conducted alongside the video data. As a result, while the original recordings contained rich multimodal information, including facial expressions, gaze shifts, gestures, and spatial positioning, these were not systematically incorporated into the transcription.

This methodological choice was deliberate and grounded in both pragmatic and conceptual considerations. The inclusion of detailed multimodal annotations often requires substantial time, expertise, and computational resources, making large-scale classroom analysis immensely complex \citep{GuerreroSosa2025}. Accordingly, this study prioritised scalability and quantitative reproducibility of verbal codes over exhaustive multimodal representation. The resulting framework thus represents a verbal-analysis-based framework conducted in conjunction with audio-video data, rather than a full-fledged multimodal learning analytics (MMLA) design. 

Despite these constraints, the AVVA framework establishes a strong methodological foundation for future multimodal extensions. Subsequent research could systematically incorporate multiple interactional modalities, such as gesture recognition, prosody and tone analysis, facial affect detection, and eye-gaze tracking to provide a richer account of learner engagement and teacher mediation. Advances in computer vision, speech recognition, and deep learning architectures can further automate the extraction of these features, enabling the large-scale integration of multimodal signals with verbal codes. Such developments would advance the framework toward a comprehensive MMLA approach, allowing researchers to triangulate verbal, visual, and behavioural data to better understand how cognitive, affective, and collaborative processes unfold in authentic classroom environments. 

In addition, future work may explore how multimodal embeddings and representation-learning techniques can link linguistic and non-verbal cues within unified analytical models. This could enhance the interpretive power of the AVVA framework by capturing not only the what of classroom discourse but also the how of interactional dynamics. Ultimately, extending verbal analysis toward multimodal integration represents a promising direction for building scalable, data-driven methodologies that maintain the interpretive richness of qualitative research while leveraging the analytical precision of modern machine learning.

\section{Summary and conclusion}
This study emphasises the continuing importance of verbal data as a central modality within the growing field of multimodal learning analytics. While technological advances now allow the capture of complex multimodal interactions through video, audio, and sensor-based data, the interpretive depth of human dialogue remains essential for understanding cognition and learning processes. The Audio Video Verbal Analysis (AVVA) framework offers a systematic approach to transform rich qualitative classroom data into quantitative representations that can be explored using statistical and computational techniques. Drawing upon \citep{chi1997} foundational work in verbal analysis and informed by contemporary multimodal research \citep{serafini2019}, the framework bridges the qualitative–quantitative divide by encoding classroom verbatims into measurable indicators for various variables. 
Triangulation is embedded as a core design principle throughout all ten steps, strengthening construct validity through convergent evidence from multiple data sources, coding perspectives, and analytical techniques. The framework acknowledges the existence of Phi Ceiling for low-frequency variables and suggests the Base Rate Filtering method, ensuring that observed associations reflect genuine co-occurrence rather than statistical distortions from rare events. Bootstrap confidence interval analysis quantifies estimation uncertainty, enabling researchers to distinguish patterns from sampling variability.In this framework, we suggest a comprehensive validation scheme that addresses fundamental challenges in temporal observational research. The Modifiable Temporal Unit Problem, in which measured associations depend on the chosen observational window, is systematically studied via a four-criterion stability assessment. Variable pairs are classified into interpretable patterns (grain-invariant, scale-specific, multi-scale) based on sign consistency, confidence interval overlap, zero exclusion, and magnitude stability across multiple temporal grain sizes. 

Application to 23 hours of classroom recordings illustrates the framework's practical viability and its capacity to produce theoretically meaningful insights. The stability analysis revealed that instructional associations manifest differently across temporal scales: certain variable pair co-occurrences remain grain-invariant (e.g., stout across all observational windows), while others exhibit scale-specific patterns (e.g., detectable only at fine or coarse temporal resolutions) or multi-scale structures (e.g., positive associations at micro-temporal scales but negative at macro-scales). These findings demonstrate that the choice of observational unit is not merely a methodological decision but a theoretical commitment about which temporal structures are under investigation. For educational practitioners and researchers, AVVA offers a pellucid, replicable methodology for systematic classroom observation that balances interpretive subtlety with analytical thoroughness. The framework's modular design allows selective adoption; researchers may apply the core coding protocol without the full validation infrastructure or implement grain-size sensitivity analysis to existing observational datasets. By clearly showing how relationships change across different time scales, the framework encourages researchers to think about why these changes happen, linking the way data are grouped to meaningful explanations about teaching and learning processes.

The Audio Video Verbal Analysis framework demonstrates that qualitative depth and quantitative rigour are not opposing methodological approaches but complementary strategies for understanding complex educational phenomena. By embedding triangulation, validation, and temporal sensitivity throughout its design, AVVA provides researchers with tools to transform classroom discourse into empirically stout, theoretically interpretable datasets. As research in education design increasingly uses different types of data and computational methods, frameworks like AVVA are important. They stay rooted in interpretive approaches while also allowing new analytical methods, helping ensure that data-driven techniques support, rather than replace, the goal of understanding teaching and learning in their full context.

% This points to apacite style and your references.bib file
\bibliographystyle{apacite}
\bibliography{references} 

\end{document}